\documentstyle[aps,preprint,floats,epsf,amsmath]{revtex}

\begin{document}
\advance\textheight by 0.2in
\draft
\title{Localization of polymers in a finite medium with fixed random obstacles }
\author{Yadin Y. Goldschmidt and Yohannes Shiferaw
\\
Department of Physics and Astronomy,
\\ 
University of Pittsburgh, Pittsburgh PA 15260, U.S.A.}
\date{\today}
\maketitle
\begin{abstract}
In this paper we investigate the conformation statistics of a Gaussian chain
embedded in a medium of finite size, in the presence of quenched random
obstacles. The similarities and differences between the case of random
obstacles and the case of a Gaussian random potential are elucidated. The
connection with the density of states of electrons in a metal with random
repulsive impurities of finite range is discussed. We also interpret the
results obtained in some previous numerical simulations.
\end{abstract}
%

\section{Introduction}

The behavior of polymer chains in random media is a well studied problem both
theoretically \cite{BM,edwards,cates,nattermann,leung,dayantis,gold,shifgold}
and experimentally \cite{cannell,rondelez,bishop,asher} and has applications
in diverse fields. For polymers, the interest arises when the chains are
confined inside an intertwined gel network \cite{asher}, and perhaps inside
porous materials and membranes \cite{cannell,rondelez,bishop}. Furthermore,
the problem is related to the statistical mechanics of a quantum particle in a
random potential \cite{gold2}, the behavior of flux lines in superconductors
in the presence of columnar defects \cite{nelson,gold3}, and the problem of
diffusion in a random catalytic environment \cite{nattermann}.

In this paper we will study the static properties of a Gaussian polymer chain,
without excluded volume interactions, that is confined in a quenched random
medium. Experimentally, a polymer in a specific solvent is known to obey
Gaussian statistics at the so called $\Theta$ temperature--where the long
range self-avoiding interactions are effectively screened. The term quenched
refers to the fact that the random medium is frozen and thus does not
thermalize with the active degrees of freedom--in this case the polymer chain.
We are interested in the properties of the polymer, such as the free energy
and radius of gyration (or alternatively the end-to-end distance), that are
averaged with the appropriate Wiener measure over all possible chain
conformations in a given realization of the random medium, with a final
average taken over all possible configurations of the random medium. The
nature of the random environment is crucial in this problem, and so it is
important to distinguish the following two important cases that have been
discussed in the literature:

1. A Gaussian random potential with short range correlations.

2. Random obstacles which prevent the chain from visiting certain sites.

\begin{flushleft}
Numerical simulations performed in three dimensions were restricted, to our
knowledge, only to the case of random obstacles
\cite{BM,leung,dayantis}. Also, a numerical
investigation based on the mapping to the Schr\"odinger equation
in one dimension was performed for the Gaussian potential case
\cite{shifgold}. On the other hand, extensive analytical work using
the replica variational approach and Flory type free energy
arguments, has been done for the case of a  Gaussian random
potential \cite{edwards,cates,nattermann,gold,shifgold}.
The case of a  bounded (saturated) random potential was also
addressed in \cite{cates}. It was not clear to us to what extent these
theoretical investigations could be applied to the case of infinitely
strong random obstacles placed randomly in the medium, as simulated numerically.
\end{flushleft}

In this paper we investigate analytically, for the first time, the case of
infinitely strong, randomly placed obstacles case. We point out
similarities and differences with the case of a Gaussian random potential, and
also the case of a saturated potential. We will assume that the obstacles are
infinitely strong--they totally exclude the chain from visiting a given site
occupied by an obstacle. Each obstacle is taken to be a block of volume
$a^{d}$, where $d$ is the number of spatial dimensions and where $a$ is the
linear dimension of the block. We take for simplicity the polymer bond length
$b$ to be approximately equal to $a$. Thus, $a$ will be the small length scale
in the problem, and we will measure all distances in units of $a$. However, in
the next section we will sometimes keep $a$ explicitly in order to omit terms
of higher order of smallness. The obstacles are placed on the sites of a cubic
lattice with lattice spacing $a$. We denote by $x$ the probability that any
given lattice site is occupied by an obstacle (block). Our main results will
concern the case of small $x$, in particular $x<x_{c}$, where $x_{c}$ refers
to the percolation threshold ($x_{c}=0.3116$ for a cubic lattice in $d=3$),
but we will also comment on the case of a larger concentration of obstacles.
We denote by ${\cal V}$ the total volume of the system.

For an uncorrelated Gaussian random potential, it was argued using qualitative
arguments in Refs. \cite{cates,nattermann} that a very long Gaussian chain
will typically curl up in some small region of low average potential. The
polymer chain is said to be localized, and for long chains the end-to-end
distance ($R_{F}$) becomes independent of chain length and scales like
\begin{equation}
R_{F}\propto(g\ln{\cal V})^{-1/(4-d)},\label{rfg}%
\end{equation}
with $g$ being the strength of the disorder (the random potential satisfies
$\langle U(x)U(x')\rangle=g \delta(x-x')$). The depth of the well entrapping the
chain is approximately
\begin{equation}
U_{min}\sim-(g\ln{\cal V})^{2/(4-d)}.\label{gsg}%
\end{equation}
These results were also obtained by the replica method in Ref. \cite{gold},
and rederived using a mapping to a quantum particle's localization in Ref.
\cite{shifgold}. For very short chains the end-to-end distance scales
diffusively ($R_{F}^{2}\sim L$), and it saturates at the $R_{F}$ value quoted
above for large $L$. Notice, that in the infinite volume limit, the chain
completely collapses. This results from the fact that the depth of the
potential is unbounded from below, and the chain is always able to find with
reasonable probability a deep enough narrow potential well to occupy,
overcoming its tendency to swell due to the entropy of confinement. The
collapse of the chain in the infinite volume limit agrees with the results for
a chain in an annealed potential, since the ability of a chain to scan all of
space for a favorable environment is equivalent to the random potential
adapting itself to the chain configuration.

To review briefly the argument leading to Eqs. (\ref{rfg}, \ref{gsg}) using
localization theory \cite{shifgold} we recall that the density of states for a
particle in an uncorrelated Gaussian random potential is given by
\cite{lifshits}
\begin{equation}
\rho(E)=\frac{A}{|E|^{\alpha}}\exp(-B|E|^{\delta}),\label{eigendist}%
\end{equation}
with $\delta=(4-d)/2$ and $B\propto1/g$, where $g$ is the strength of the
disorder. This result is valid for an infinite volume. In a finite volume
${\cal V}$ the energy will be bounded from below. We can estimate the
lowest energy $E_{c}$ from the tail of the distribution:
\begin{equation}
\int_{-\infty}^{E_{c}}dE\rho(E)\simeq1/{\cal V},
\end{equation}
which leads to
\begin{equation}
E_{c}=-\left(  \frac{\ln{\cal V}}{B}\right)  ^{1/\delta}.
\end{equation}
The width of the ground state wave function (localization length) is given by
\begin{equation}
\ell_{c}\sim|E_{c}|^{-1/2}\sim\left(  g\ln{\cal V}\right)  ^{-1/(2\delta)}
\ .
\end{equation}
The mapping from a quantum particle of mass $m$, at a finite temperature
$1/\beta$, to a polymer chain, is given by
\begin{equation}
\hbar\rightarrow T,\ \ \hbar\beta\rightarrow L,\ m\rightarrow dT/b^{2}
\ .\label{mapping}%
\end{equation}
It can be shown that the ground state width $\ell_{c}$ is proportional to the
end-to-end distance $R_{F}$ of a chain that is situated in a deep minimum
in the volume ${\cal V}$, whose depth is given by $E_{c} \sim U_{min}$.
Using the above mapping the density matrix for a quantum particle at finite
temperature \cite{feynman} corresponds to the partition sum (Green's function)
of a Gaussian polymer chain \cite{doi}.

\section{Random obstacles}

For the case of infinitely strong randomly placed obstacles, the potential
energy of the polymer chain is always zero. Hence, the free energy of the
polymer will be $F=-TS$, since $E=0$, and the statistics of the polymer will
be dictated only by entropic effects. As the volume of the system tends to
infinity there is a chance to find very large lacunae free of obstacles. Thus,
in the limit of large volume we do not expect the polymer to collapse, but
rather to inflate with increasing number of monomers ($L$). We will now
analyze the behavior of a polymer in an environment consisting of random
obstacles and find that there are three different phases as a function of the
volume of the system. In the subsequent analysis we will always assume that
$L$ is very large.

In order to estimate the average chain properties we first assume that a very
long polymer chain will attain an approximately spherical shape of radius $R$
(we will discuss this spherical droplet approximation in the sequel). Now, let
us coarse grain the volume ${\cal V}$ into subregions of volume $v \sim
R^{d}$, and assume that the polymer is confined to one of these regions. Each
of these coarse grained subregions will contain a different number of
obstacles, and the chain will reside in that region with the lowest number of
obstacles which can be found in the finite volume ${\cal V}$ in order to
minimize its free energy. We will estimate $R$ by writing the free energy of
the polymer as a function of both $R$ and the coarse grained volume fraction
of obstacles in a given subregion (which also depends on $R$), and minimizing
it accordingly. First, let us assume that there are no obstacles present
inside this spherical region. The entropy of a chain confined in this
``cavity'' is of the form
\[
S=L\ln(z)-\alpha\frac{L}{R^{2}},
\]
where $z$ is the number of nearest neighbors, e.g. $2d$ for a cubic lattice in
$d$ dimensions, and the second term is the entropy of confinement
\cite{degennes}. Here, $\alpha$ is a numerical constant. The free energy will
then be $F=-TS$. In the following we choose $T=1$ for simplicity since the
temperature does not play any significant role with respect to the results.

In order to proceed and estimate the entropy change due to obstacles inside
the volume $v$, we use the mapping from the polymer to a quantum particle as
discussed in the introduction. The free energy per monomer of the polymer when
$L$ is very large corresponds to the ground state energy of a quantum particle
in a cavity of radius $R$. This is known, in three dimensions, to be equal to
$E_{0}=(\hbar^{2}/2m)\pi^{2}/R^{2}$, in agreement with the expression above
(up to an unimportant additive constant). In $d$ dimensions the energy is
still proportional to $1/R^{2}$ but the prefactor is different. Suppose that
there is a spherical obstacle of radius $a$ inside the sphere. If the obstacle
is at the center of the sphere the Schr\"{o}dinger equation is exactly
solvable and the ground state in $d=3$ is given by
\begin{equation}
\Psi(r)=C\frac{\sin\frac{\pi(r-R)}{R-a}}{r},\ \ a<r<R
\end{equation}
and $\Psi(r)=0$ otherwise. This will correspond to an energy of
\begin{equation}
\frac{m}{\hbar^{2}}E_{0}=\frac{\pi^{2}}{2R^{2}}+\frac{\pi^{2}a}{R^{3}}+\ldots,
\end{equation}
where the corrections vanish faster than $a$ as $a\rightarrow0$. If on the
other hand the obstacle is not in the center of the sphere we could only find
an approximate solution of the Schr\"odinger equation which can be used to
give an upper bound to the ground state, which may be exact to leading order
in $a$ (see Appendix). The ground state energy becomes
\begin{equation}
\frac{m}{\hbar^{2}}E_{0}=\frac{\pi^{2}}{2R^{2}}+\frac{\pi^{2}a}{R^{3}}\left(
\frac{R}{\pi R_{0}}\sin\frac{\pi R_{0}}{R}\right)  ^{2}+\ldots,
\end{equation}
where $R_{0}$ is the distance of the center of the obstacle from the center of
the sphere. One can see that the factor in parenthesis approaches $1$ as
$R_{0}\rightarrow0$, and vanishes as $R_{0}\rightarrow R$. Notice, that for
the analysis above we have treated the obstacles as spherical in shape as
opposed to a square block. However, this difference should only amount to an
unimportant numerical factor.

Using the mapping from the quantum particle to the polymer as given by
Eq.~(\ref{mapping}), we find
\begin{equation}
\frac{m}{\hbar^{2}}E_{0}\rightarrow\ \frac{3F}{a^{2}L},
\end{equation}
where $F$ is the free energy of the chain. We have used the fact that $T=1$
and $b=a$ for the bond length. (This correspondence is up to an
additive constant that is given by the entropy of a chain in free
space). Now, if the position of the obstacle is random
we expect that we should average the energy over all the possible locations of
the obstacle within the volume of the sphere to obtain
\begin{equation}
\frac{3F}{a^{2}L}=\frac{\pi^{2}}{2R^{2}}+\frac{3a}{2R^{3}}+\ldots
.\label{en1obst}%
\end{equation}
Let us denote by $\hat{x}$ the volume fraction occupied by the random
obstacles within the spherical volume $v$. Then the number of obstacles inside
this volume will be $\frac{4\pi}{3}(R/a)^{3}\hat{x}$. If $\hat{x}$ is small
the energy due to several obstacles will be approximately equal to the sum of
the individual energies. The deviation from this rule becomes important only
if the obstacles are very close or touching each other, and for small $\hat{x}$ the number
of such configurations is very small compared with configurations where the
obstacles don't touch. In any case interactions among obstacles are at
least of O($\hat{x}^2$). Thus, the free energy of a polymer chain confined to a
volume $v$, with a volume fraction $\hat{x}$ occupied by obstacles, is
\begin{equation}
\frac{F}{L}=\frac{\pi^{2}}{6(R/a)^{2}}+\frac{2\pi}{3}\hat{x} +\ldots.\label{enx}%
\end{equation}
The important conclusion is that the term proportional to $\hat{x}$ is
independent of $R$. The numerical prefactors are really of no importance to
us. The small $\hat{x}$ approximation is enhanced by the fact that not only
do we
assume that $x$ is small ($x<x_c$), but the chain will be found to settle in regions
where the obstacle concentration is smaller than average. In two dimensions we
find that the first term in Eq.~(\ref{en1obst}) is proportional to $1/R^{2}$
and the second term to $1/(R^{2}|\ln(a/R)|)$ (see Appendix). In one dimension
the situation is totally different since an obstacle will divide the volume
$v$ into two disjoint regions. Thus, all our conclusions apply only
above two dimensions, where the second term in Eq.~(\ref{en1obst}) is
proportional to $a^{d-2}R^{-d}$. When multiplying by the number of obstacles
$\sim(R/a)^{d} \hat{x}$ one gets an $R$-independent result for the second term
in Eq.~(\ref{enx}). Two dimensions is a borderline case where some subtleties
may arise. In the following we will measure all distances in units of $a$ so
we put $a=1$.

We now proceed to study the chain statistics by considering the fluctuations
of the volume fraction $\hat{x}$. If $x$ is the probability that a lattice
site is occupied by an obstacle, then the coarse grained volume fraction
$\hat{x}$ is distributed according to the binomial probability distribution
$b(v\hat{x};v,x)$. The notation
\begin{equation}
b(k;n,p)=\binom{n}{k}p^{k}q^{n-k},\label{binom}%
\end{equation}
stands for the probability that $n$ Bernoulli trials with probabilities $p$
for success and $q=1-p$ for failure result in $k$ successes and $n-k$ failures
\cite{feller}. For $\hat{x}=0$ it gives the so called Lifshits probability to
find an empty region of volume $v$ free of obstacles. 
With this probability
there will be associated an ``entropy'' which will be its logarithm.

Using the results above we can start to discuss the statistics of a polymer in
an infinite volume (${\cal V}$ $\rightarrow\infty$)--the so called
annealed result. Assuming that the chain takes on a roughly spherical
configuration of volume $v\sim R^{d}$, the free energy will read
\begin{equation}
F(R,\hat{x})=-L\ln(z)+\frac{L}{R^{2}}+L\hat{x}-\ln[b(R^d\hat{x};R^d,x)].
\label{fullfree}%
\end{equation}
This free
energy has to be minimized both with respect to $\hat{x}$, and to $R$, since
the chain is free to move and find the most favorable values for these
parameters. The most favorable value of $\hat{x}$ for large $L$ and for an
infinite volume is $\hat{x}=0$, since $\hat{x}$ is not allowed to be
negative. Using the fact that
\begin{eqnarray}
b(0;v,x)=(1-x)^v,
\label{bin0}
\end{eqnarray}
we find that the expression for the free energy becomes
\begin{equation}
F(R)=-L\ln(z)+\frac{L}{R^{2}}-R^{d}\ln(1-x).
\label{free0}
\end{equation}
This free energy has now to be minimized with respect to $R$ to yield
\begin{equation}
R_{m,annealed}\sim\left(  \frac{L}{|\ln(1-x)|}\right)  ^{1/(d+2)}%
.\label{annealedR}%
\end{equation}
Thus the size of the chain grows with $L$, but with an exponent
smaller than $1/2$, the free chain exponent.

So far we discussed the case of an infinite volume ${\cal V}$. In a finite
volume we find that the so called quenched and annealed case differ, at least
when the volume is not too big. We actually find that there are three regions
as a function of the size of the system volume ${\cal V}$. First, if
${\cal V<V}_{1}\simeq\exp(x^{-(d-2)/2}/(1-x))$, it is unlikely for a chain
of volume $v\sim R^{d}$ to find a region which is totally free of
obstacles. Thus $\hat{x}$ does not vanish in this regime. To proceed further we must
use an approximation to the binomial distribution $b(v\hat{x};v,x)$.

If $v$ is not too small
we can approximate the binomial distribution by a normal distribution \cite{feller}
\begin{equation}
b(v\hat{x};v,x)\approx(2\pi vx(1-x))^{-1/2}\exp\left(  -\frac{v(\hat{x}%
-x)^{2}}{2x(1-x)}\right)  .
\end{equation}
This approximation is good provided $vx\gg1$ and $v(1-x)\gg1$. We will verify
below that these conditions are indeed met in our case when $x$ is small.

In a finite volume ${\cal V}$, the lowest expected value of $\hat{x}$, to be
denoted by $\hat{x}_{m}$, can be found from the tail of the distribution
\begin{equation}
\int_{0}^{\hat{x}_{m}}d\hat{x}\exp\left(  -\frac{v(\hat{x}-x)^{2}}%
{2xy}\right)  \simeq\frac{v}{{\cal V}},
\label{xm}
\end{equation}
which gives
\begin{equation}
\hat{x}_{m}\simeq x-\sqrt{\frac{xy\ln{\cal V}}{R^{d}}}.
\end{equation}
The free energy becomes
\begin{equation}
F_{I}(R)=-L\ln(z)+\frac{L}{R^{2}}+Lx-L\sqrt{\frac{xy\ln{\cal V}}{R^{d}}.}%
\end{equation}
The last term in Eq.~(\ref{fullfree}) is missing since it is negligible for
large $L$ when $R$ is independent of $L$. Minimizing $F(R)$ with respect to $R$ we find
\begin{equation}
R_{mI}\sim\left(  xy\ln{\cal V}\right)  ^{-1/(4-d)}\label{Rm1}%
\end{equation}
and%
\begin{equation}
\hat{x}_{mI}=x-\left(  xy\ln{\cal V}\right)  ^{2/(4-d)},
\label{xmI}
\end{equation}
where we put $y=1-x$.
The result for the radius of gyration of the chain, as represented by $R_{mI}$
is the same result as for the case of the Gaussian distributed random
potential, but with the strength $g$ replaced by $x(1-x)$. The polymer in this
case is localized and its size is independent of $L$ for large $L$. 

As ${\cal V}$ grows $R_{m}$ decreases until eventually $\hat{x}_{m}$ vanishes.
This happens when ${\cal V=V}_{1}\simeq\exp(x^{-(d-2)/2}y^{-1})$. For
${\cal V>V}_{1}$, $R_{m}$ is no longer given by Eq.~(\ref{Rm1}), but rather
by the solution of $\hat{x}_{mII}=0$. It is the largest region free of
obstacles expected to be found in a volume ${\cal V}$. Rather than
using the normal approximation we can estimate $R_m$ directly from the relation
\begin{eqnarray}
(1-x)^v\simeq v/{\cal V},
\end{eqnarray}
with $v\sim R_m^d$. Solving for $R_m$ we obtain
\begin{eqnarray}
R_{mII}\sim\left(  \frac{\ln{\cal V}}{|\ln(1-x)|}\right)^{1/d}.
\label{Rm2}
\end{eqnarray}
The polymer is still localized but the dependence on $x$ and on $\ln
{\cal V}$ has changed. In this region which we call region II the free
energy is given by
\begin{equation}
F_{II}=-L\ln(z)+\alpha L\left(  \frac{\ln{\cal V}}{|\ln(1-x)|}\right)
^{-2/d},\label{F2}%
\end{equation}
where some undetermined constant $\alpha$ is introduced for later convenience.

As ${\cal V}$ grows in region II, $R_{mII}$ continues to grow until it
reaches the annealed value given in Eq.~(\ref{annealedR}). This happens when
\begin{equation}
{\cal V=V}_{2}\sim\exp\left(  x^{2/(d+2)}L^{d/(d+2)}\right)
\end{equation}
to leading order in $x$, which is enormous for large $L$. For ${\cal V>V}%
_{2}$ we have the third region in which $R_{mIII}=R_{m,annealed}$ and it grows
like $L^{1/(d+2)}$.

Since in region I we have used the normal approximation to the
binomial distribution, we should check \textit{a posteriori} if the condition
$vx\gg1$ is met. Since ${\cal V<V}_{1}$ we find
\begin{equation}
C_{d}R_{mI}^{d}x\approx C_{d}x^{-(d-2)/2}\left(  \frac{\ln{\cal V}_{1}}%
{\ln{\cal V}}\right)  ^{d/(4-d)}>C_{d}x^{-(d-2)/2}.
\end{equation}
Here $C_{d}=2\pi^{d/2}/(d\Gamma(d/2))$ is the volume of a $d$-dimensional
sphere of unit radius. In $d=3$, for example, $C_{3}=4\pi/3$. 
Thus we see that for $d>2$, $vx\gg1$ provided $x\ll1$.
The minimum value is attained for ${\cal V=V}_{1}$, and is significantly
larger when ${\cal V<V}_{1}$. It is interesting to notice that in $d=3$ for
example, even if $x=0.3$, we find that $C_{3}/\sqrt{x}=7.6$ which is
considered large enough for the validity of the normal approximation ( a value
of 5 is usually considered sufficient).

Actually, the  normal approximation to the binomial distribution
$b(k;n,p)$ is not entirely acccurate even if the conditions $np\gg1$ and
$nq\gg1$ are met, if $k$ is far from the center, i.e. if $|k-np|^3/n^2>1$
\cite{feller}. In our case we need that $v|\hat{x}_m-x|^3<1$ to be
satisfied. In region I we have
\begin{eqnarray}
C_d R_{mI}^d |\hat{x}_{mI}-x|^3<C_d (xy\ln{\cal V}_1)^{(6-d/(4-d)}=C_d
x^{(6-d)/2},  
\end{eqnarray}
where we have used Eq.~(\ref{xmI}) and the value of ${\cal V}_1$.
For $d=3$ this is smaller than one even for $x=0.3$. 

The behavior in region II can also be deduced from known results of the
density of states for a quantum particle in the presence of obstacles
(repulsive impurities). In that case \cite{lifshits} the density of states is
given by (when the obstacles are placed on a lattice)
\begin{equation}
\rho(E)\sim\exp(-c|\ln(1-x)|E^{-d/2}),\ E>0
\end{equation}
with $c$ being some dimension dependent constant and $x$ is the density of
impurities. Note that $\rho(E)$ vanishes for $E<0$. We can estimate the lowest
energy in a finite volume ${\cal V}$ from the integral
\begin{equation}
\int_{0}^{E_{c}}dE\rho(E)\simeq1/{\cal V},
\end{equation}
and find
\begin{equation}
E_{c}\sim\left(  \frac{\ln{\cal V}}{|\ln(1-x)|}\right)^{-2/d},
\end{equation}
and thus the localization length is given by
\begin{equation}
\ell_{c}\sim|E_{c}|^{-1/2}\sim\left(  \frac{\ln{\cal V}}{|\ln(1-x)|}\right)  ^{1/d},
\end{equation}
which coincides with $R_{mII}$ above. 

To conclude this section we make some remarks on the validity of the spherical
droplet approximation. The shape of a long polymer chain is determined by the
regions of the random medium that have a lower than average number of
obstacles. For ${\cal V}>{\cal V}_{1}$ these regions are essentially
free of obstacles. The probability of finding such empty regions depends only
on its volume and not its shape. However given regions of varying shapes and
equal volumes, it will be entropically more favorable for a long polymer chain
to reside in a region whose shape is closest to a sphere. This is because the
confinement entropy is maximized for a sphere over other shapes of the same
volume. The argument is equivalent to that proposed by Lifshits
\cite{lifshits} in the context of electron localization and is shown rigorously
by Luttinger \cite{luttinger}. For ${\cal V}>{\cal V}_{1}$ the relevant
regions contain a small number of obstacles but we believe that the same
argument should roughly hold and deviations from a spherical shape will be small
or irrelevant.

\section{Comparison with numerical simulations}

We have compared our results from the last section with numerical simulations
performed by Dayantis \textit{et al. }\cite{dayantis}, and also comment on the
relation to earlier simulations done by Baumgartner and Muthukumar \cite{BM}.
Dayantis \textit{et al. } carried out simulations of free chains
(random-flight walks) confined to cubes of various linear dimensions $6-20$, in
units of the lattice constant. These chains can intersect freely and lie on a
cubic lattice. They introduced random obstacles with concentrations $x=0, 0.1,
0.2$ and $0.3$. The length of the chains vary between $18-98$ steps. They also
simulated self-avoiding chains that we will not discuss here. They measured
the quenched entropy, the end-to-end distance, and also the radius of gyration
which is a closely related quantity. Unfortunately, these authors did not have
a theoretical framework to analyze their data, and thus could not make it
collapse in any meaningful way. We show below how it is possible to fit the
data nicely to our analytical results.

Even for $x=0.1$, the value that we get for ${\cal V}_{1}$ is about $33$
which is an order of magnitude smaller than the the smallest volume used in
their simulation, which is $216$ for a cube of side $6$. Hence we expect to be
in region II. To check the agreement with our analytical results we show in
Figure 1 a plot of $\ln(-S/L+\ln6)$ vs. $\ln(\ln{\cal V}/|\ln(1-x)|)$ where $S$ is
the entropy measured in the simulations and ${\cal V}=B^{3}$ for a box of
side $B$. Recall that $F=-S$ and Eq.~(\ref{F2}) predicts a straight line with
slope $-2/3$. The best fit is obtained for a slope of $-0.72\pm0.05$, which is
in excellent agreement with our analytical results in region II.

\begin{figure}[ptb]
\centerline{\epsfysize8.5cm \epsfbox{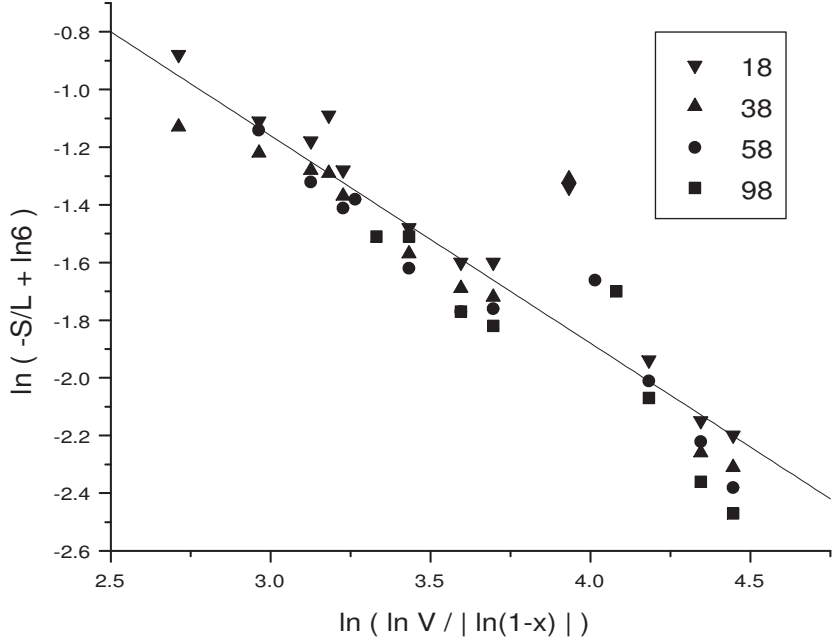}} \vspace{3mm}\caption{A plot
of $\ln(-S/L+\ln6)$ vs. $\ln(\ln{\cal V}/|\ln(1-x)|)$. The labels are
marked according to the chain length.}%
\label{figure1}%
\end{figure}

In order to analyze the simulation results for the end-to-end distance and
radius of gyration we have to introduce some additional compensation for the
results obtained in the previous section. First we must realize that
Eq.~(\ref{Rm2}) is valid only when the number of steps (monomers) is very
large. In the simulations they used chains of varying lengths whose size did
not yet reach asymptotia. Hence, we introduce a correction factor
\begin{equation}
R_{m}(L)=R_{m}(1-\exp(-L/R_{m}^{2}))^{1/2}\equiv R_{m}f_{1}(\sqrt{L}/R_{m})
\ ,
\end{equation}
which interpolates between the size of a free chain as $L\rightarrow0$ and the
value of $R_{m}$ from Eq.~(\ref{Rm2}) as $L\rightarrow\infty$.

The second correction we have to implement arises when the expected value of
the chain is not much smaller than the size of the confining box. Even for a
free chain confined to a box of side $B$ with no obstacles present, the
end-to-end distance is not simply $R=L^{1/2}$ for $L^{1/2}<B$ and $R=B$ for
larger $L$. We have to take into account the fact that the length of the chain
has a Gaussian distribution about its expected value, and the tail of the
Gaussian is cut off by the presence of the box (this is for the absorbing
boundary conditions that is used in the simulations). Thus, for the case of no
obstacles ($x=0$), The measured end-to-end distance should approximately be
\begin{equation}
R_{c}^{2}=\int_{-B}^{B}dR\ R^{2}\exp(-\frac{R^{2}}{2L})/\int_{-B}^{B}%
dR\ \exp(-\frac{R^{2}}{2L}),
\end{equation}
which gives $R_{c}=\sqrt{L}f_{2}(B/\sqrt{L})$ with
\begin{equation}
f_{2}(x)=\left(  1-\sqrt{\frac{2}{\pi}}\frac{x}{\operatorname{erf}(x/\sqrt
{2})}\exp(-x^{2}/2)\right)  ^{1/2}.
\end{equation}
This indeed gives good agreement with the measured values in the no obstacle
case. For the obstacle case we thus have to introduce these two corrections in
subsequent order:
\begin{equation}
R_{m,corrected}=R_{m}f_{1}(\sqrt{L}/R_{m})f_{2}\left(  \frac{B}{R_{m}%
f_{1}(\sqrt{L}/R_{m})}\right)  ,\label{Rcalculated}%
\end{equation}
where $R_{m}=R_{mII}$ as given by Eq.~(\ref{Rm2}). (A constant of
proportionality of $1.8$ has been introduced on the rhs of
Eq.~(\ref{Rm2}) to obtain a good fit).
In Figures 2 and 3 we show
a comparison of the simulation results for the end-to-end distance and for the
radius of gyration with the calculated results as given by Eq.~(\ref{Rm2}) and
Eq.~(\ref{Rcalculated}). The agreement seems remarkable, especially for
the end-to-end distance, where all the data collapses to a straight
line with a slope close to $1$.

\begin{figure}[ptb]
\centerline{\epsfysize8.5cm \epsfbox{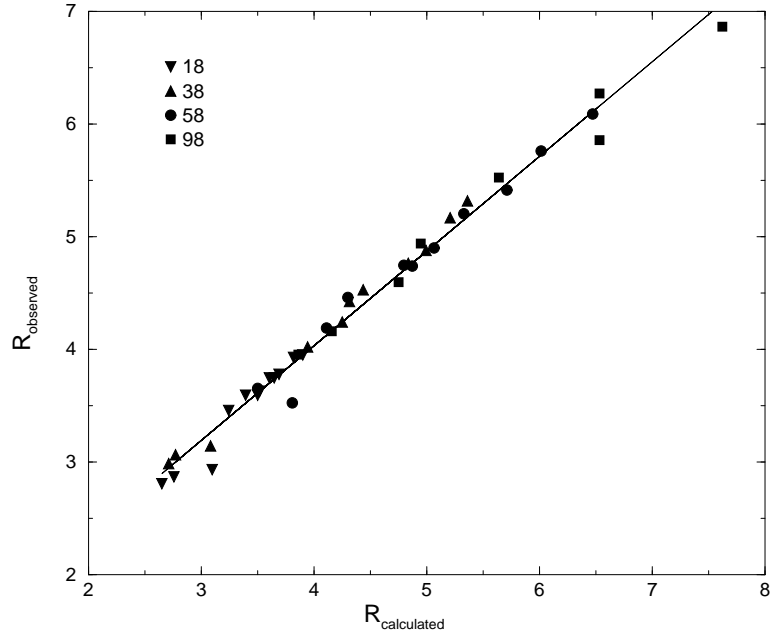}} \vspace{3mm}\caption{A plot of
the observed vs. calculated end-to-end distance}%
\label{figure21}%
\end{figure}

\begin{figure}[ptb]
\centerline{\epsfysize8.5cm \epsfbox{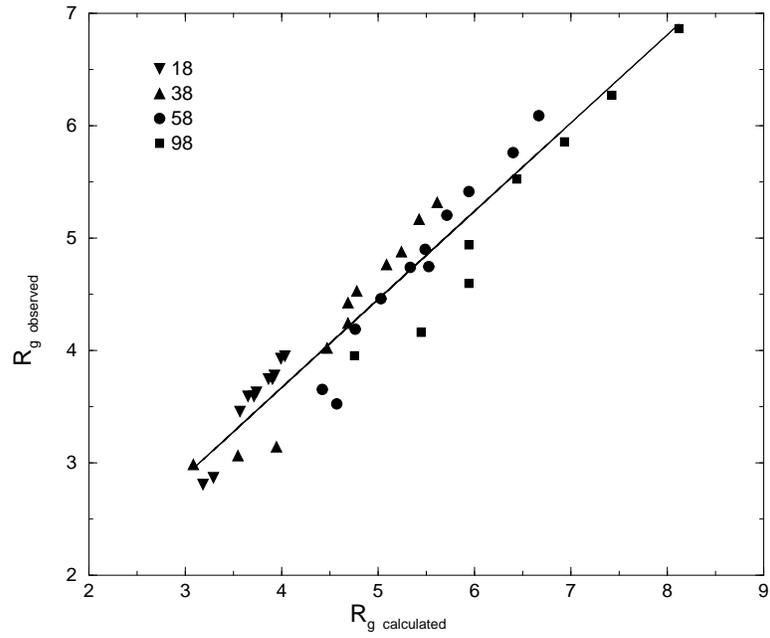}} \vspace{3mm}\caption{A plot of
the observed vs. calculated radius of gyration}%
\label{figure3}%
\end{figure}

Dayantis \textit{et al.} emphasize that they did not consider concentrations
of obstacles above the percolation threshold, which is at $x_{c}=0.3116$ for a
simple cubic lattice. The reason is that above the percolation threshold the
medium of random obstacles begins to form disconnected islands free of
obstacles. Thus, in their simulation the polymer chain will only sample a
limited fraction of the volume available. What happens is that effectively the
volume available for the chain is not the total volume of the cube but rather
the volume of the disconnected region it occupies. For most realizations of
the random medium this effective volume will be smaller than the value
${\cal V}_{1}$, which is the limit of region I of the last section. In that
case one expects the end-to-end distance to scale like $x^{-1}$ as given in
Eq.~(\ref{Rm1}) instead of like $x^{-1/3}$ as given by Eq.~(\ref{Rm2}).
Baumgartner and Muthukumar's simulation was for both below the percolation
threshold and also above it ($x=0.4$ and $0.5$). However, they only estimate
the exponent above the percolation threshold, and find it to be about $-1$.
They do not estimate the exponent for $x$ below the percolation threshold,
which appears from their data to scale with a much smaller exponent. Thus, it
seems likely that the reason these authors report a behavior corresponding to
region I, even though their box is quite large, is because the effective
volume is small for the cases for which they exceed the percolation threshold.

\bigskip

\section{Summary and Discussion}

In this paper we have considered the effect of random obstacles on the
behavior of a free Gaussian chain. We have seen that in the presence of
infinitely strong obstacles, that exclude the chain from visiting randomly
chosen sites, there are three possible behaviors of the end-to-end distance as
a function of $L$ (the number of monomers). The three possible regimes depend
on the total volume of the system. If the volume of the system is smaller than
${\cal V}_{1}\simeq\exp(x^{-(d-2)/2})$, where $x$ is the probability that a
lattice site is occupied, then the polymer is localized, as in the case of a
Gaussian random potential, and $R_{I}\sim\left(x\ln{\cal V}\right)
^{-1/(4-d)}$. For ${\cal V}_{1}<{\cal V}<{\cal V}_{2}$, where ${\cal V}%
_{2}\sim\exp\left(  x^{2/(d+2)}L^{d/(d+2)}\right)  $, the polymer size is
given by $R_{II}\sim\left(x/\ln{\cal V}\right)^{-1/d}$. Finally, for
${\cal V>V}_{2}$, the polymer behaves the same way as for an annealed
potential, i.e. $R_{III}\sim\left(L/x\right)^{1/(d+2)}$. We have
displayed the results to leading order in $x$ for small $x$. These results
are valid only when the average volume fraction of the obstacles ($x$) is smaller than the
percolation threshold. When $x$ is bigger than the percolation
threshold we expect that the system breaks into independent domains whose
volume is independent of, and generally much smaller than, the total volume of the system.

We were able to fit nicely the simulations of Dayantis \textit{et al.}
\cite{dayantis} to the analytic behavior in region II. The results of Baumgartner and
Muthukumar for $x>x_{c}$, presumably correspond to region I. Region
III is very difficult to be seen in simulations since ${\cal V}_{2}$ is huge
for large $L$. However such behavior, which coincides with the annealed case
have been obtained by Chandler \textit{et al.} \cite{leung} in their
simulations of annealed random obstacles.

The behavior in the three regions has some similarities to the case of a
saturated random potential as discussed by Cates and Ball \cite{cates}.
However, there are also significant differences. The saturated potential
discussed by these authors concerned the case where the random potential can
take two different values $\pm\sqrt{g}$ with probability $1/2$ each. They
considered the case when $g$ is small. In that case, when the chain occupies
an approximately spherical volume $R^{d}$, its energy is estimated by the
average potential in that region times the number of monomers $L$ of the
chain. This leads to a coarse grained renormalized potential with a Gaussian
probability distribution that is valid provided the predicted size of the
chain is much larger than the largest cavity that is free of obstacles. On the
other hand, in the case of infinitely strong obstacles the free energy of a
chain occupying a volume $R^{d}$ has a completely entropic origin. The energy
of the chain remains zero as it is completely excluded from the regions
occupied by obstacles. In the case of a saturated potential the behavior in
region I is given by $R_{SI}\sim\left(  g\ln{\cal V}\right)  ^{-1/(4-d)}$,
so in this region $g$ plays a similar role to $x$ (the subscript $S$ refers to
the saturated potential case) . However, the volume of the system beyond which
this behavior is no longer valid is given by ${\cal V} _{S1}\simeq
\exp(g^{-d/4})$, which is quite different from its value in the strong
obstacle case. Here $d=2$ is no longer a lower critical dimension, and the
result, which has a very different $d$-dependence, is valid down to (and
including) one dimension. For ${\cal V}_{S1}{\cal <V<V}_{S2}$, one has
$R_{SII}\sim\left(  \ln{\cal V}\right)  ^{1/d}$, independent of $g$. Also
${\cal V}_{S2}\sim\exp\left(  L^{d/(d+2)}\right)  $, independent of $g$.
Above ${\cal V}_{S2}$ the annealed result $R_{SIII}\sim L^{1/(d+2)}$ applies.

As compared to the unsaturated gaussian random potential, the two models
coincide only in region I when ${\cal V}<{\cal V}_{1}$. For ${\cal V}
\rightarrow\infty$ the polymer collapses to a point in the Gaussian potential
case as it can find a very deep and narrow potential well, whereas in the
random obstacle case the polymer swells as $L^{1/(d+2)}$ with growing $L$ as
it can find large lacunae free of obstacles. Our analysis of the simulations
of Dayantis \textit{et al.} \cite{dayantis} shows clearly that their results do not
conform to the behavior dictated by the Gaussian model but are described well
by the expected behavior in the intermediate region II.
\begin{acknowledgments}
This research is supported by the US Department of Energy (DOE), grant No.
DE-FG02-98ER45686.
\end{acknowledgments}
%

%
\appendix

\section*{}

Consider a spherical cavity of radius $R$ about the origin. Inside the cavity
there is one obstacle of radius $a$, centered at location ${\bf R}_{0} $.
The potential within the obstacle is infinite and thus the wave function has
to vanish in that region. We will assume that $a\ll R$. Let us define the
unperturbed wave function
\begin{equation}
\Psi_{0}({\bf r})=\frac{A}{|{\bf r}|}\sin\frac{\pi|{\bf r}|}%
{R}.\label{psi0}%
\end{equation}
This is the ground state solution of the Schr\"{o}dinger equation in the absence of the obstacle.

Consider now the following trial wave function:
\begin{equation}
\Psi({\bf r})=\left\{
\begin{array}
[c]{c}%
0,\ \ \ \ \ \hspace{1in}\hspace{1in}\ \ \ \ \ 0<|{\bf r}-{\bf R}%
_{0}|<a\\
\Psi_{0}({\bf r})\Psi_{1}({\bf r-R}_{0}),\ \hspace{1in}%
\ \ \ \ \ a<|{\bf r}-{\bf R}_{0}|<r_{m}\\
\Psi_{0}({\bf r}),\ \hspace{1in}\ \ \ \ \ \ \ \ \ \ \ r_{m}<|{\bf r}%
-{\bf R}_{0}|\ \ \&\&\ \ |{\bf r}|<R
\end{array}
\right. \label{psi}%
\end{equation}
with
\begin{equation}
\Psi_{1}({\bf r})=\frac{r_{m}}{\sin\left(  \frac{\pi(r_{m}-a)}{R}\right)
}\frac{1}{|{\bf r}|}\sin\left(  \frac{\pi(|{\bf r}|-a)}{R}\right)
.\label{psi1}%
\end{equation}
Here $r_{m}$ is determined by the condition
\begin{equation}
{\bf \nabla}\Psi_{1}({\bf r})|_{|{\bf r}|=r_{m}}=0,\label{seam}%
\end{equation}
which leads to the condition
\begin{equation}
\tan\frac{\pi(r_{m}-a)}{R}=\frac{\pi r_{m}}{R}.
\end{equation}
In the limit of small $a$ this leads to
\begin{equation}
r_{m}\simeq\left(  \frac{3}{\pi^{2}}\right)  ^{1/3}R^{2/3}a^{1/3}.\label{rm}%
\end{equation}
Thus the volume of the region around the obstacle in which the wave function
deviates from $\Psi_{0}$ is about $4R^{2}a/\pi$, which vanishes like $a$ as
$a\rightarrow0$ , not like $a^{3}$. In the following we will refer to this
volume as $V_{a}$. The choice of $r_{m}$ is designed to insure that not only
the wave function but also its gradient are continuous across the seam
$|{\bf r}-{\bf R}_{0}|=r_{m}$. Thus As $|{\bf r -R}_{0}|\rightarrow
r_{m}$ from below
\begin{align}
{\bf \nabla}\Psi=\Psi_{1} {\bf \nabla}\Psi_{0}+\Psi_{0} {\bf \nabla
}\Psi_{1} \rightarrow{\bf \nabla}\Psi_{0},
\end{align}
since ${\bf \nabla}\Psi_{1} \rightarrow0$ and $\Psi_{1} \rightarrow1$.

Using this wave function an upper bound on the energy is given by
\begin{equation}
\frac{m}{\hbar^{2}}E_{0}\leq\frac{\int\Psi({\bf r})(-\frac{1}%
{2}{\bf \nabla}^{2})\Psi({\bf r})dV}{\int\Psi^{2}({\bf r})dV}.
\end{equation}
Evaluating the rhs we find
\begin{equation}
\frac{m}{\hbar^{2}}E_{0}\leq\frac{\pi^{2}}{2R^{2}}+\frac{\int\Psi
({\bf r})(-\frac{1}{2}{\bf \nabla}^{2}-\frac{\pi^{2}}{2R^{2}}%
)\Psi({\bf r})dV_{a}}{\int\Psi_{0}^{2}({\bf r})dV}+O(a^{2}%
),\label{shift}%
\end{equation}
where in the numerator we only integrate over the volume $V_{a}$ about the
obstacle. Now, inside $V_{a}$%
\begin{equation}
{\bf \nabla}^{2}\Psi({\bf r})=\Psi_{1}({\bf r-R}_{0}){\bf \nabla
}^{2}\Psi_{0}({\bf r})+\Psi_{0}({\bf r}){\bf \nabla}^{2}\Psi
_{1}({\bf r-R}_{0})+{\bf \nabla}\Psi_{0}({\bf r})\cdot{\bf \nabla
}\Psi_{1}({\bf r-R}_{0}).\label{lappsi}%
\end{equation}
We argue that
\[
\int dV_{a}{\bf \nabla}\Psi_{0}({\bf r})\cdot{\bf \nabla}\Psi
_{1}({\bf r-R}_{0}),
\]
vanishes faster than $a$ in the limit $a\rightarrow0$, because of the angular
integration: ${\bf \nabla}\Psi_{0}({\bf r})$ points in a fixed direction
away from the origin at ${\bf 0}$, whereas ${\bf \nabla}\Psi
_{1}({\bf r-R}_{0})$ points in a varying radial direction away from the
center at ${\bf R}_{0}$, and thus the angular integration will yield
$\int_{-1}^{1}\cos\theta\ d\cos\theta=0$. We have actually verified that the
contribution of this term is O($a^{5/3}$), hence negligible. Since $-(1/2)$
${\bf \nabla}^{2}\Psi_{0}({\bf r})=(\pi^{2}/2R^{2})\Psi_{0}({\bf r}%
)$, the first term in Eq.~(\ref{lappsi}) exactly cancels the second term of
the integral in the numerator of Eq.~(\ref{shift}). Thus we are left with
\begin{equation}
\frac{m}{\hbar^{2}}E_{0}\leq\frac{\pi^{2}}{2R^{2}}+\frac{\int\Psi_{0}%
^{2}({\bf r})\Psi_{1}({\bf r-R}_{0})(-\frac{1}{2}{\bf \nabla}%
^{2})\Psi_{1}({\bf r-R}_{0})dV_{a}}{\int\Psi_{0}^{2}({\bf r})dV}.
\end{equation}
At this point we can approximate $\Psi_{0}^{2}({\bf r})$ by its value at
${\bf r=R}_{0}$, and take out of the integral. Also,
\[
(-\frac{1}{2}{\bf \nabla}^{2})\Psi_{1}({\bf r-R}_{0})=(\pi^{2}%
/2R^{2})\Psi_{1}({\bf r-R}_{0}).
\]
Thus
\begin{equation}
\frac{m}{\hbar^{2}}E_{0}\leq\frac{\pi^{2}}{2R^{2}}+\frac{\pi^{2}}{2R^{2}}%
\Psi_{0}^{2}({\bf R}_{0})\frac{\int\Psi_{1}^{2}({\bf r)}dV_{a}}{\int
\Psi_{0}^{2}({\bf r})dV}.
\end{equation}
Evaluating the integrals and using the estimate for $r_{m}$ as given by
Eq.~(\ref{rm}), we arrive at the final result
\begin{equation}
\frac{m}{\hbar^{2}}E_{0}=\frac{\pi^{2}}{2R^{2}}+\frac{\pi^{2}a}{R^{3}}\left(
\frac{R}{\pi R_{0}}\sin\frac{\pi R_{0}}{R}\right)  ^{2}+\ldots,
\end{equation}
as given in section II, where $R_{0}$ is the magnitude of ${\bf R}_{0}$.
Although we have used an approximate wave function that can give only an upper
bound on the correction, it appears likely that the correction might be exact
to leading order in $a$. Support for this comes from the fact that it coincides
with the exact answer in the limit $R_{0}=0$.

We now consider briefly the two dimensional case. In this case we consider
only the case where the obstacle is in the middle of the circular region.
Without the obstacle present, the ground state solution to the Schr\"{o}dinger equation is
given by
\begin{equation}
\Psi_{0}({\bf r})=A{\mathrm J}_{0}\left(  \frac{2.405r}{R}\right)  ,
\end{equation}
where J$_{0}$ is the spherical Bessel function and $x_{0}\simeq$2.405 is its
first zero. This leads to
\begin{equation}
\frac{m}{\hbar^{2}}E_{0}=\frac{1}{2}\frac{x_{0}^{2}}{R^{2}}.
\end{equation}
With an obstacle of radius $r$ present at the center, we look for a wave
function of the form
\begin{equation}
\Psi_{0}({\bf r})=\left\{
\begin{array}
[c]{c}%
A{\mathrm J}_{0}\left(  kr\right)  +B{\mathrm N}_{0}\left(  kr\right)
,\hspace{0.5in}r>a\\
0,\hspace{0.5in}\hspace{0.5in}\hspace{0.2in}\hspace{0.2in}\hspace
{0.2in}\ \ r<a
\end{array}
\right.
\end{equation}
where ${\mathrm N}_{0}$ is Neumann's function, sometimes denoted by Y$_{0}$.
The variable $k$ is determined by the requirement that $\Psi_{0}(r)=0$ at $r=a$ and $r=R$.
This leads to the condition
\begin{equation}
\frac{{\mathrm J}_{0}\left(  kR\right)  }{{\mathrm N}_{0}\left(  kR\right)
}=\frac{{\mathrm J}_{0}\left(  ka\right)  }{{\mathrm N}_{0}\left(  ka\right)  }.
\end{equation}
Thus
\begin{equation}
\frac{{\mathrm J}_{0}\left(  kR\right)  }{{\mathrm N}_{0}\left(  kR\right)
}\approx\frac{1+O(k^{2}a^{2})}{(2/\pi)(\ln(ka)+\gamma-\ln2)+O(k^{2}a^{2})},
\end{equation}
and we find
\begin{equation}
k\approx\frac{x_{0}}{R}+\frac{\alpha}{R|\ln\frac{a}{R}|},
\end{equation}
with
\begin{equation}
\alpha=\frac{\pi}{2}\frac{{\mathrm N}_{0}(x_{0})}{{\mathrm J}_{0}^{\prime
}\left(  x_{0}\right)  }.
\end{equation}
Here J$_{0}^{\prime}$ is the derivative of J$_{0}$ with respect to its
argument. It thus follows that
\begin{equation}
\frac{m}{\hbar^{2}}E_{0}\approx\frac{1}{2}\frac{x_{0}^{2}}{R^{2}}+\frac
{x_{0}\alpha}{R^{2}|\ln\frac{a}{R}|}.
\end{equation}

\end{document}